\documentclass[12pt,preprint]{aastex}
\newcommand{\gtwid}{\mathrel{\raise.3ex\hbox{$>$\kern-.75em\lower1ex\hbox{$\sim$}}}}
\newcommand{\ltwid}{\mathrel{\raise.3ex\hbox{$<$\kern-.75em\lower1ex\hbox{$\sim$}}}}

\shorttitle{Thermal Emission as a Test for Hidden AGN}
\shortauthors{Whysong and Antonucci}

\begin{document}

\title{Thermal Emission as a Test for Hidden Nuclei in Nearby Radio Galaxies}

\author{D. Whysong and R. Antonucci}
\affil{Physics Department, University of California, Santa Barbara, CA 93106}

\begin{abstract}
It is widely believed that the optical/UV continuum of quasars (the "Big Blue
Bump") represents optically thick thermal emission from accretion onto a black
hole. Narrow line radio galaxies don't show such a component directly, and
were historically thought for that reason to be rotation-powered, with large
kinetic luminosity in the radio jets but very little accretion or optical
radiation. When the Unified Model came along, identifying at least some narrow
line radio galaxies as hidden quasars, the compelling observational motivation
for this radio galaxy scenario lost some of its force. However, it is far from
clear that all narrow line radio galaxies contain hidden quasar nuclei.

The clear sign of a hidden quasar inside a radio galaxy is the appearance
of quasar spectral features in its polarized (scattered) light. However that
observational test requires suitably placed scattering material to act as a
mirror, allowing us to see the nuclear light. A rather robust and more general
test for a hidden quasar is to look for the predicted high mid-IR luminosity
from the nuclear obscuring matter.

The nuclear waste heat is detected and well isolated in the nearest narrow line
radio galaxy, Cen A. This confirms other indications that Cen A {\em does}
contain a modest quasar-like nucleus. However we show here that M87 {\em does
not}: at high spatial resolution, the mid-IR nucleus is seen to be very weak,
and consistent with simple synchrotron emission from the base of the radio jet.
This fairly robustly establishes that there are "real" narrow line radio
galaxies, without the putative accretion power, and with essentially all the
luminosity in kinetic form.

Next we show the intriguing mid-IR morphology of Cygnus A, reported previously
by us and later discussed in detail by Radomski et al.\ (2002). All of this
mid-IR emission is consistent with reprocessing by a hidden quasar, known
to exist from spectropolarimetry by Ogle et al.\ (1997) and other evidence.

\end{abstract}

Subject headings: galaxies: active --- galaxies: individual (M87, Cen A, 3C
405, 3C 273) --- infrared: galaxies

\keywords{quasar, narrow line radio galaxy}

\section {Introduction}

Now that the existence of supermassive black holes in AGN seems fairly
secure, perhaps the next most fundamental questions are the sources of energy
and the nature of the accretion flow in the various classes of objects.
Historically (see Begelman, Blandford and Rees 1984 for an early review) it was
thought that the optical/UV continuum (or ``Big Blue Bump") in quasars
(hereafter: and broad line radio galaxies) represents thermal radiation from
some sort of cool optically thick accretion flow. Radio galaxies didn't show
this component, so were posited to be ``nonthermal AGN" with hot radiatively
inefficient accretion, at a very low rate; in these cases the the jet power
would derive from the hole rotation rather than release of gravitational
potential energy from accretion.

These arguments took a surprising turn when it was realized that many radio
galaxies do have the quasar-like nuclei\footnotemark[1] that are invisible from
our line of sight. One Fanaroff-Riley II (edge-brightened, very luminous)
radio galaxy, 3C234, was shown in 1982 to have quasar features (broad permitted
emission lines and Big Blue Bump) in polarized light (Antonucci 1982, 1984;
Antonucci 1993 for a review). Thus 3C234 does have the ``thermal" optical/UV
emission, which is only visible via scattering. Many other examples have been
shown subsequently (e.g.\ Hines and Wills 1993; Young et al.\ 1996). Some
invocations of the Unified Model postulated that this was generally true of the
FR II (powerful, edge-brightened) class (e.g. Barthel 1989).

3C234 has powerful high-ionization narrow lines, consistent with its being a
hidden quasar. However it is still contentious how those FR IIs with weak
and/or low ionization narrow emission lines fit in (Singal 1993; Laing 1994;
Gopal-Krishna et al.\ 1996; Antonucci 2001). The situation is even less clear
for the FR I galaxies, almost all of which have undetectable or low-ionization
emission lines.

For radio galaxies with no observable high-ionization narrow emission line
region present, there is no {\em a priori}\/ evidence for the presence of a
quasar. These could still have a quasar nucleus, but any narrow line region
would need to be mostly obscured as well. In fact there is some evidence that
the high ionization narrow emission lines are partially obscured in many 3C
radio galaxies (Hes et al.\ 1993). In NGC 4945 and many other ordinary looking
galaxies, the only present evidence for a hidden AGN is in the hard X-ray
(e.g. Madejski et al.\ 2002).

To summarize, {\em many}\/ FR II radio galaxies fall into the apparently
weak, low-ionization emission line category. {\em Most}\/ of the FR I radio
galaxies do as well. These spectra have been described as "optically dull."
Quite recently, Chiaberge et al.\ (1999, 2000) have shown that {\em among these
optically dull sources of both FR types}, a majority show optical point sources
in HST images, and of course more might do so with better imaging data. Note
that their result seems surprising at first: one might have guessed that those
with stronger high-ionization spectra would have the point sources, but the
opposite is true.

Chiaberge et al.\ argue that those with detectable optical point sources cannot
in general have thick obscuring tori on. The reason is, {\em if} the point
sources are truly nuclear, then we can see to the very center in most of these
objects. These radio galaxies are effectively selected by an isotropic emission
property - the radio lobe flux - so the source orientations should have an
isotropic distribution. That is, the radio axes should be randomly oriented
with respect to our line of sight. Since we can see optical point sources in
most optically dull objects, most lines of sight to their nuclei must be
unobscured in general. Recall that in the Chiaberge et al.\ picture, the HST
optical point sources are the bases of the synchrotron jets which emit in the
radio. As they point out, one caveat must be given with their line of argument.
It is possible that obscuring tori exist below the HST resolution, and hide the
nucleus and the very innermost region of the conical jet. For example, their
optical point sources may really be jet emission on $\sim 1$ pc scales,
and optical Big Blue Bump sources are much smaller. A torus might be large
enough to obscure the latter while still allowing the pc scale jet emission
to be seen over the top.

In the Chiaberge et al.\ scenario the optical point sources represent
synchrotron emission from the bases of the jets. In support of this idea,
those authors show that the optical fluxes correlate roughly with the core
radio fluxes at 6\ cm. It should be easy to check this: the relationship
should tighten up substantially using millimeter fluxes instead of those at
6\ cm. 

It is very important to remember that some narrow line radio galaxies of {\em both}
FR types do have strong high ionization emission lines, and in some cases
definitive spectropolarimetric evidence for a hidden "thermal" nucleus. In
general such objects have no detectable optical point source. They behave
instead like Seyfert 2 nuclei, with just spatially resolved scattered light
and the extended narrow line region directly visible.

Our goal is to determine robustly which if any radio galaxies {\em lack} a
hidden "thermal" (optical/UV continuum) nucleus. This is crucial for AGN
theory since it would prove the existence of an accretion mode different from
that in quasars. In particular, current wisdom would posit a very low
accretion rate and a very low radiative efficiency, that is, some variant of the
advection dominated accretion flow for those radio galaxies. Then by default
the enormous kinetic luminosity of the radio jets would be attributed to black
hole rotational energy.

How can we tell whether or not a hidden nucleus is present? One method is
via the hard X-rays. Many hidden AGN that are not Compton-thick have been
discovered with X-ray observations. Some references to penetrating X-rays in
optically dull sources are gathered in Antonucci 2001; NGC4945 is a spectacular
example (e.g. Madejski et al.\ 2000).

Our approach here is to look instead for reradiation of the absorbed light
from any hidden quasar-like nucleus by the dusty obscuring matter (torus).
Modulo factors of order unity, the various models of the obscuring tori predict
that the "waste heat" from the obscuring matter will emerge in the mid-IR, and
that this emission is {\em roughly isotropic} in all but the highest
inclination cases. (Note that this assumption is the weakest point of our
project, but that the various torus models predict it is true {\em to within a
factor of a few}: Pier and Krolik 1992, 1993; Granato et al.\ 1997;
Efstathiou and Rowan-Robinson 1995; a related model in Konigl and Kartje 1994
should also be consulted; Cen A is discussed specifically in Alexander et al
1999). We are making this test for hidden quasars using the Keck I telescope's
mid-IR instrument (the "LWS"). This is the only way to isolate the nucleus
well, and to achieve the required sensitivity. Here we show the power of the
technique, and demonstrate the existence of at least one "nonthermal"
AGN\footnotemark[2].

\section {Observations}

\subsection{3C 405 (Cygnus A)}

in Figure 1 we present a diffraction limited mid-IR image of the nuclear source
in 3C 405 (Cyg A), obtained with the Long Wavelength Spectrometer (LWS)
instrument at the Keck I telescope\footnotemark[3]. All data were taken with
the 11.7$\mu$m filter, which has an $\sim1\ \mu$m bandpass from 11.2 to 12.2
$\mu$m. The image was shown previously in Whysong and Antonucci (2001).

The nucleus of 3C 405 was imaged at 11.7$\mu$m with Keck I/LWS on 1999 September
30. The chop/nod throw was set to 10{\tt "} in order to allow imaging
of larger scale extended structure; this places the chop beam off
the chip, which has a 10{\tt "} field. We do not report on structures larger
than the 10{\tt "} chop distance. The images were dithered in a 5 position
box pattern, 2{\tt "} to a side, with 53.1 seconds on-source for the positive
image per dither position. The entire 5-position exposure was repeated three
times, for a total on-source time of 796.6 seconds.

Data were processed by subtracting all background chop/nod frames, shifting
each dithered image to the correct position, and coadding all dither images.
Morphology is extended, with structure to the east and southeast of the nucleus
(Fig.~1).  The standard star was alpha Ari\footnotemark[4], with a FWHM of
0.27{\tt "}.

\notetoeditor{Table 2 belongs here}

Our 11.7$\mu$m image and a partial spectral energy distribution are shown
in Figs.~1 and 5. For comparison, the IRAS (large aperture) data for Cygnus A
are listed in Table 3.

\notetoeditor{Table 3 belongs here}


\subsection {Cen A}

The high resolution mid-IR data on Cen A also come from the Keck I telescope.
Data were obtained on our behalf by Randy Campbell on 28 June 2002. Three
filters were used: the 11.7 $\mu$m filter with 1 $\mu$m bandpass, a wider
"SiC" filter centered at 11.7 $\mu$m but with a 2.4 $\mu$m bandwidth, and
the 17.75 $\mu$m filter with a $\sim 1\ \mu$m bandwidth. The standard star
was Sigma Sco, a multiple star which was partially resolved so that it was
necessary to increase the synthetic aperture to 2.24 arcsec in order to include
all components. No photometric data are available for Sigma Sco at wavelengths
longer than M, so we extrapolated to longer wavelengths using the
Rayleigh-Jeans approximation.

The Cen A images show only an unresolved source in all filters. Photometric
calibration results are F($11.7 \mu$m)$=1.6$ Jy, F(SiC)$=1.8$ Jy with $\approx
0.3$ arcsec FWHM resolution, and F(17.75\ $\mu$m)$=2.3$ Jy with $\approx 0.5$
arcsec FWHM resolution.

Since Cen A is so close, there are published data for relatively small physical
apertures. A small physical aperture is key, because for both Cen A and M87,
the large-aperture (e.g. IRAS, ISO) fluxes are much larger than that from the
nucleus. But the much higher resolution Keck data isolate a nuclear point
source with size $\approx 5$ pc at $11.7 \mu$m and $\approx 7.8$ pc at $17.75
\mu$m given a distance of 3.1 Mpc.

We will assume that this is mostly dust emission heated by the nucleus. This is
consistent with the 10-20 micron nuclear spectrum taken by ISOCAM CVF in a
$\sim 4"$ aperture (Mirabel et al.\ 1999, reproduced here as Fig. 3). In fact,
our 11.7 $\mu$m flux is in good agreement with that of the spectrum, suggesting
that Fig. 3 is in fact the nuclear spectrum.

For comparison, the published, larger aperture mid-IR fluxes are much higher.
The Cen A central region flux has been measured in the mid-IR by Grasdelen and
Joyce (1976). Their 3.5" aperture has the same flux as the 5.2" aperture, so
there is a compact source surrounded by a region of low or zero flux.  The 3.5"
aperture corresponds to $~$50\ pc, and the enclosed flux is given as $\sim 2.6$
Jy at 11 microns, and 4.3 Jy at 12.6 microns. The IRAS 12.6$\mu$m flux measurement
is even higher at 13.3 Jy.

\subsection {M87}

Here a new high-resolution Keck I image is also crucial. We obtained this
data for M87 on 2000 January 18. The observation was made in chop-nod mode
using a small 3.5{\tt "} amplitude so as to keep both the object and chop beams
on the CCD chip. The 11.7 $\mu$m filter was used, with a 1$\mu$m bandpass.

Integration time was 96 seconds per dither each for positive and negative
(background).  Unfortunately, due to a loss of guiding, the positive nucleus
image was only fully imaged in one dither frame. However, we are still left
with two independent images, both strong pointlike detections. Our measurement
was taken from the better placed of the two.

Beta And and Mu UMa were used as standards for photometric calibration,
yielding a flux scale of 0.0874 and 0.0931 mJy/(ADU/s) and FWHM of 0.33{\tt "}
and 0.31{\tt "} respectively. This calibration for the unresolved nuclear
component in M87 results in a flux of 13 +/- 2 mJy. The uncertainty is
dominated by systematic errors in the background subtraction; we adopt a value
of 13 mJy. A synthetic aperture of 0.96 arcsec was used, but the source is
unresolved so the flux is insensitive to the aperture.

Again we note that the large aperture fluxes are much higher.  In particular
the IRAS fluxes (Moshir et al.\ 1990) are 231$+/-$37 mJy at 12$\mu$m,
$<$241 mJy at 25$\mu$m, and 393$+/-$51 mJy at 60$\mu$m.


\section {Discussion}

\subsection {Cen A}

Fig 1 shows the SED for Cen A, combining all the $\sim1$" measurements.  For
AGN which show the quasar-like optical nucleus in scattered light only, the
mid-IR emission is typically two orders of magnitude greater than the
(scattered) optical light. For Cen A the situation is similar, but with a
slight modification (Bailey et al.\ 1996; Hough et al.\ 1987; Antonucci and
Barvainis 1990; Alexander et al.\ 1999). The ratio is inflated by absorption of
the {\em pc} scale scattered optical light by the kpc dust lane famous from
photographs.

At 2u, a highly polarized point source indicates detection of a $\sim$\ pc scale
reflection region.  The near-IR reflected light from pc scales penetrates the
{\em kpc} scale dust lane which has only moderate optical depth. This situation
was deduced for 3C323.1 and Cen A by Antonucci and Barvainis (1990). The
pc scale dust screen can be consistently identified with the cold absorber seen
in the X-ray spectrum, which has column density $\sim 3 \times 10^{23}$ \
cm$^{-2}$. Additional optical and near-IR scattered light on kpc scales has
been mapped and disucssed by Capetti et al 2000 and Marconi et al 2000.

The case of Cen A shows that even radio galaxies with weak or low-ionization
lines (the ``optically dull" ones) can have hidden Type 1 nuclei. Others can
be found in e.g. Ekers and Simkin 1983 and Sambruna et al.\ 2000.

To show the level of the reflected light in the nuclear region, we have
plotted the value at 2 microns rather than that in the optical (which is
highly absorbed).  The K-band nuclear flux with starlight subtracted (Marconi
et al.\ 2000) is F(K) $\sim$ 3 mJy, and $\nu L_{\nu} \sim 6 \times 10^{39}$
ergs/sec for a distance of 3 Mpc.  The SED in Figure 2 shows that the mid-IR
luminosity of the nucleus is much larger than the optical/near-IR value, as
expeted. This is consistent with all other reflected-light objects.


The value of the mid-IR flux quoted above is $\sim$3.5 Jy at 11.7 microns,
interpolated from measurements at two nearby wavelengths.  The corresponding
$\nu L_{\nu}$ is $1.2 \times 10^{42}$ erg/sec.  Assuming a normal quasar SED
(e.g.  Sanders et al 1989; Barvainis et al.\ 1990), that translates to a
bolometric luminosity of $~2 \times 10^{43}$ ergs/sec for the hidden
AGN. 

The nature of the mid-IR emission is important. The SED has been fit to a
synchrotron self-Compton model (Chiaberge et al.\ 2001) which would lead to
a classification for Cen A as a ``misaligned BL Lac". However, the fit was to
the mid-IR continuum slope instead of the different ISO bands because the
latter were heavily affected by absorption and emission features (M. Chiaberge
2001, private communication). Small aperture mid-IR spectra show PAH and Si
features (Figure 3), and the SED in $\leq 4"$ apertures is consistent with
predominantly dust rather than synchrotron (Alexander et al.\ 1999;
note also that the ISOCAM CVF flux from the Mirabel et al.\ 1999 spectrum is
consistent with our 11.7 $\mu$m measurement, suggesting that the spectrum is
representative of our smaller aperture). These features indicate that most of
the mid-IR flux is thermal emission as expected for the torus model.

\subsection {M87}

This radio galaxy is on the FR I-II borderline, both in morphology and in radio
power (Owen et al.\ 2000).  It is one of the majority of FR I radio galaxies
with an optical/UV point source (Chiaberge et al.\ 1999, 2000). The optical
point source is tentatively ascribed to synchrotron radiation associated with
the radio core (see Ford \& Tsvetanov 1999, as well as the Chiaberge et al.\
papers).  However it's not known whether this light is highly polarized, or
whether broad emission lines are strong in either total or polarized optical/UV
flux.

Our small $\sim0.3$ arcsec FWHM beam isolates the innermost $\sim25$pc in M87.
The enclosed 11.7$\mu$m flux in an 0.6" synthetic aperture is 15 mJy; this
aperture matches that used for Cen A in physical size.  The flux measured in
this way is much lower than the large aperture measurements in the literature.
Published large-aperture data leave plenty of room for waste heat from a hidden
AGN, but our data do not (Fig.~1).  The mid-IR luminosity $\nu L_{\nu}$ is only
of order that in the optical rather than much greater as for the hidden AGN
sources.  In fact, much or all of the mid-IR flux could simply be synchrotron
radiation associated with the innermost part of the jet, so the measured flux
is an upper limit to the dust luminosity. Note that this outcome for M87 is
just what Chiaberge et al.\ implicitly predicted.

Unless the nuclear dust is too obscured to emit in the mid-IR, this rules out a
powerful hidden nucleus.  The observed 11.7$\mu$m flux corresponds to $\nu
L_{\nu} = 1.0\times 10^{41}$ erg/sec, for a distance of 15 Mpc.  Suppose the
mid-IR core is in fact all dust emission.  For the SED for the PG quasar
composite of Sanders et al.\ 1989, a bolometric luminosity of $\sim1.6\times
10^{42}$ ergs/sec is expected.  For comparison, a lower limit to the jet
kinetic luminosity in M87 is $\sim5\times10^{44}$ erg/sec (Owen et al.\ 2000),
so the jet is by far the dominant channel for energy release.\footnotemark[6]
If correct, this suggests that M87 is the true ``misaligned BL Lac."

Published ADAF models (Reynolds et al.\ 1996) predict very low IR-optical-UV
luminosities compared with those in the radio and X-ray.  Our 11.7$\mu$m point
is about equal to the optical value, but certainly our 11.7$\mu$m
point may be partially or completely jet emission\footnotemark[7].

\subsection{Cygnus A}

This is a very powerful FR II radio galaxy at a redshift of 0.056.  It has
strong high-ionization narrow lines, suggestive of a hidden AGN.  A broad Mg II
2800 emission line is detectable in total flux (Antonucci et al.\ 1994).  That
line may or may not be highly polarized, and thus scattered from a hidden
nucleus. Several detailed papers report spectroscopic and spectropolarimatric
data (Goodrich and Miller 1989; Tadhunter et al.\ 1994; Shaw and Tadhunter
1994; Vestergaard and Barthel 1993; Stockton et al.\ 1994; see also Tadhunter
et al.\ 2000 and Thornton et al.\ 1999 - and there are several others),
culminating in Ogle et al.\ 1997, which shows an extremely broad H-alpha line
in polarized flux. It is virtually invisible in total flux because its great
width makes it hard to distinguish from continuum emission.

A nuclear point source in the near-IR was noted by Djorgovski et al.\ (1991),
but they don't seem to have considered hot dust emission for this excess over
the extrapolation from the optical light, as we believe.

A powerful hidden nucleus should manifest a mid-IR dust luminosity much larger
than the observed optical luminosity. However for this object and M87 (and
virtually all others!) the only IR data available were taken with very large
beam sizes. We (and Radomski et al.\ 2002)\footnotemark[8] isolate the core much
better with the $\sim0.3\ \rm arcsec$ ($\sim 1.1$ kpc) resolution provided by
the Keck I telescope, and find a nuclear flux of $\sim 60$mJy. An uncertainty
here derives from the extended emission, but flux as a function of aperture
size does flatten out for apertures larger than the seeing disk, so the
0.96{\tt "} measurement should be approximately correct (see Table 2). However,
we can't be sure from this observation alone that the emission is on pc scales.
Since the emission is powerful and at the relatively short wavelength of
11.7$\mu$m, it is very likely that this comes from nuclear dust rather than a
starburst.

It is entirely possible that the extended emission is thermal dust even at
radii up to 1 kpc. The temperature of nuclear-heated dust can be estimated
according to Barvainis 1988. Adopting a Hubble constant of 75 km/s Mpc, the
$\nu L_{\nu}$ luminosity at 11.7$\mu$m is $9.2\times 10^{43}$ erg/sec.  If the
intrinsic SED is similar to those of PG quasars (Sanders et al.\ 1989), the
11.7$\mu$m value implies a bolometric luminosity of 16.5 $\nu L_{\nu}$
(11.7$\mu$m) $\sim 1.5\times 10^{45}$ erg/sec. An optical luminosity of
$\approx 10^{45}$ erg/s produces a dust temperature of $\approx 120\ $K at a
500pc radius.  A similar calculation of the dust temperature was done by
(Radomski et al.\ 2002), yielding a slightly higher results (150 K) due to a
higher estimate of the optical luminosity. Among the uncertainties are the
nuclear UV luminosity and the possiblility of single photon heating.

The IRAS (large aperture, see Table 3) dust spectrum is quite cool, suggesting
a large starburst contribution. Extended emission is seen in our image at
11.7$\mu$m. The core can't be exactly separated from the extensions (see Table
2), but we can estimate around 60 mJy for the nuclear dust. Fig.~1 shows the
Cyg A 11.7$\mu$m image, and Fig.~5 shows a partial spectral energy
distribution. The nuclear luminosity $\nu L_{\nu}$ at 11.7$\mu$m is 10 times
higher than that at 0.5$\mu m$. The latter wavelength needs two roughly
canceling corrections: subtraction of optical light from the host galaxy, and
dereddening (Ogle et al.\ 1997). The starbust contribution to the nuclear
11.7$\mu$m emission is expected to be small, but this should be checked with
a spectral slope measurement.

The conclusion is simple and expected from prior evidence: Cyg A has a
moderately powerful hidden nucleus.

As noted above, the estimated bolometric luminosity is $1.5 \times 10^{45}$ erg/s.
For comparison the jet power is estimated several different ways (Carilli and
Barthel 1996; Sikora 2001; Punsly 2001). The values are rough, but generally
lie in the $\gtwid10^{45}$ ergs/sec range. This is consistent with the finding
that jet power and optical/UV luminosity are often comparable in double radio
quasars (e.g., Falcke 1995).

Thus this is a moderate luminosity broad line radio galaxy with a very high
luminosity radio source. It has in fact been inferred already that Cygnus A is
an over-achiever in the radio (Carilli and Barthel 1996; Barthel and Arnaud
1996). This is well explained qualitatively by the fact that it is the only
known FR II radio source in a rich X-ray-emitting cluster.

\subsection {Blazars}

A goal of this paper is to detect thermal IR bumps, often in the presence of a
synchrotron continuum. In general the SED sampling is limited for our targets.
Even in cases with convincing dust emission, the dust component is not very
well isolated in the infrared SED. Thus we thought it would be helpful to show
a well observed object which clearly isolates a nonthermal infrared bump from
a broadband nonthermal component.

Blazar radio sources are typically dominated over most of their SED by variable
synchrotron emission (and/or inverse Compton bump in the X-ray - $\gamma$-ray
region). This means that in powerful cases, relatively few continuum components
of emission are important for the SED. For example, starlight and dust warmed
to tens of K by stars are relatively weak. Thus the infrared consists almost
entirely of the broadband nonthermal and the (relatively) narrow band thermal
components. The SED, shown in Fig. 4, is that of 3C 273 (data from Robson et
al.\ 1993).  It can be referred to while assessing the other SEDs in this
paper.

\section{Relation to other Radio Galaxies and Conclusion}

Radio lobe emission is fairly isotropic, so it's easy to make lists of double
radio sources that are nearly unbiased with respect to orientation. The
visibility of optical point sources in {\em most}\/ of the optically dull (weak
or low ionization emission line) galaxies show that there are no $\gtwid$
parsec scale tori present which are able to obscure the unresolved optical
sources in those cases (Chiaberge et al.\ 1999, 2000). Whatever the nature of
the M87 optical point source, we know it's small from the variability on year
timescales. The same would apply to the other optically dull nuclei if they
vary as M87 does.  This would strengthen the Chiaberge et al.\ argument that we
have unobscured sightlines nearly all the way to the central engines.

Since the M87 optical/UV flux is quite variable (e.g., Tsvetanov et al.\ 1998)
jet synchrotron emission is a possibility. By correlating the radio
synchrotron core fluxes and the optical point source fluxes in FR I radio
galaxies generally, Chiaberge et al.\ (1999) infer that the latter are in fact
likely to be beamed synchrotron sources. Crucial tests of the nature of the
optical point sources can be made with spectroscopy and polarimetry. We hope to
do this with adaptive optics, excluding most of the starlight that dominates in
arcsec apertures.

However, a large minority of low ionization FR I and FR II radio galaxies show
no point source, and are similar in this way to AGN with hidden nuclei. In fact
the closest FR I, Cen A, {\em does}\/ have a big molecular torus (see Fig.~2 of
Rydbeck et al.\ 1993!), and substantial evidence for a hidden nucleus as well
(see references cited earlier). As a working hypothesis we might suppose that
the same is true for all those without detectable optical pointlike nuclei. (Of
course sensitivity of the optical/UV observations also enters in.)

Thus the FR I family is heterogeneous: some contain hidden optical/UV nuclei
and some do not. It's been difficult to find FR I objects with strong evidence
for hidden AGN, but in fact at least a few are known to be quasar-like from
direct spectroscopy (3C120 is well known; see also Lara et al.\ 1999 and
Sarazin et al.\ 1999). The infrared peaks in some blazars such as 3C 66A are
clearly distinct from the radio synchrotron, providing further evidence for
FR I objects with hidden thermal emission. Therefore the ``nonthermal" model
{\em does not} apply to all optically dull or low ionization radio galaxies, or
to all FR I galaxies. The FR class has no apparent {\em direct} relation to the
mode of energy production, consistent with much recent evidence that FR Is
behave very much like FR IIs at VLBI scales. We will understand this better
after the completion of our mid-IR program.

Finally we reiterate an assumption in our program: we assume that any
mid-IR dust emission is isotropic to within a factor of a few.

\section{Acknowledgements}

We wish to thank M. Chiaberge, P. Ogle, B. Wills, and R. Barvainis for
good advice. A thousand huzzas to Randy Campbell at Keck for obtaining
the Cen A data.

\footnotetext[1]{We will use the terms "thermal AGN" and "hidden AGN" to refer
to the optical/UV continuum and broad emission lines which characterize the
spectra of quiescent quasars, broad line radio galaxies, and Seyfert galaxies
in this spectral region.}

\footnotetext[2]{Preliminary reports of this work were published in Anotnucci
2001 and Whysong and Antonucci 2001 (astro-ph 0106381); shortly thereafter,
Gemini mid-IR images were published and analyzed by Perlman et al.\ 2001. Their
measurements and conclusions were similar to ours. Since their images were very
deep, they were also able to detect extended jet emission.}

\footnotetext[3]{Instrument reference is available at:
http://www2.keck.hawaii.edu:3636/realpublic/inst/lws/lws.html.}

\footnotetext[4]{A table of photometric standards is available at:
http://www2.keck.hawaii.edu:3636/realpublic/inst/lws/IRTF$\_$Standards.html.}


\footnotetext[6]{We can also make the empirical test of the ratio of the mid-IR
flux to that of the radio lobes. We measured 15mJy at $11.7\mu$m, and NED
shows $\sim220$Jy at 1.4GHz. The ratio is $6.8 \times 10^{-5}$, much lower
than the ratio for the objects with hidden AGN, such as Cen A.}

\footnotetext[7]{It is unclear to us why the radio points, which fit the ADAF
model, are taken as measurements while the (non-fitting) optical fluxes were
not. Also, the Reynolds et al.\ figure apparently uses 3C273 as a ``thermal"
quasar-like template, but that object definitely has a large jet contribution
in the radio and infrared (Robson et al.\ 1993).}

\footnotetext[8]{Our Keck I image appeared publicly before that of Radomski et
al. 2002, but we did relatively little analysis.}



\clearpage

\begin{figure}
\plotone{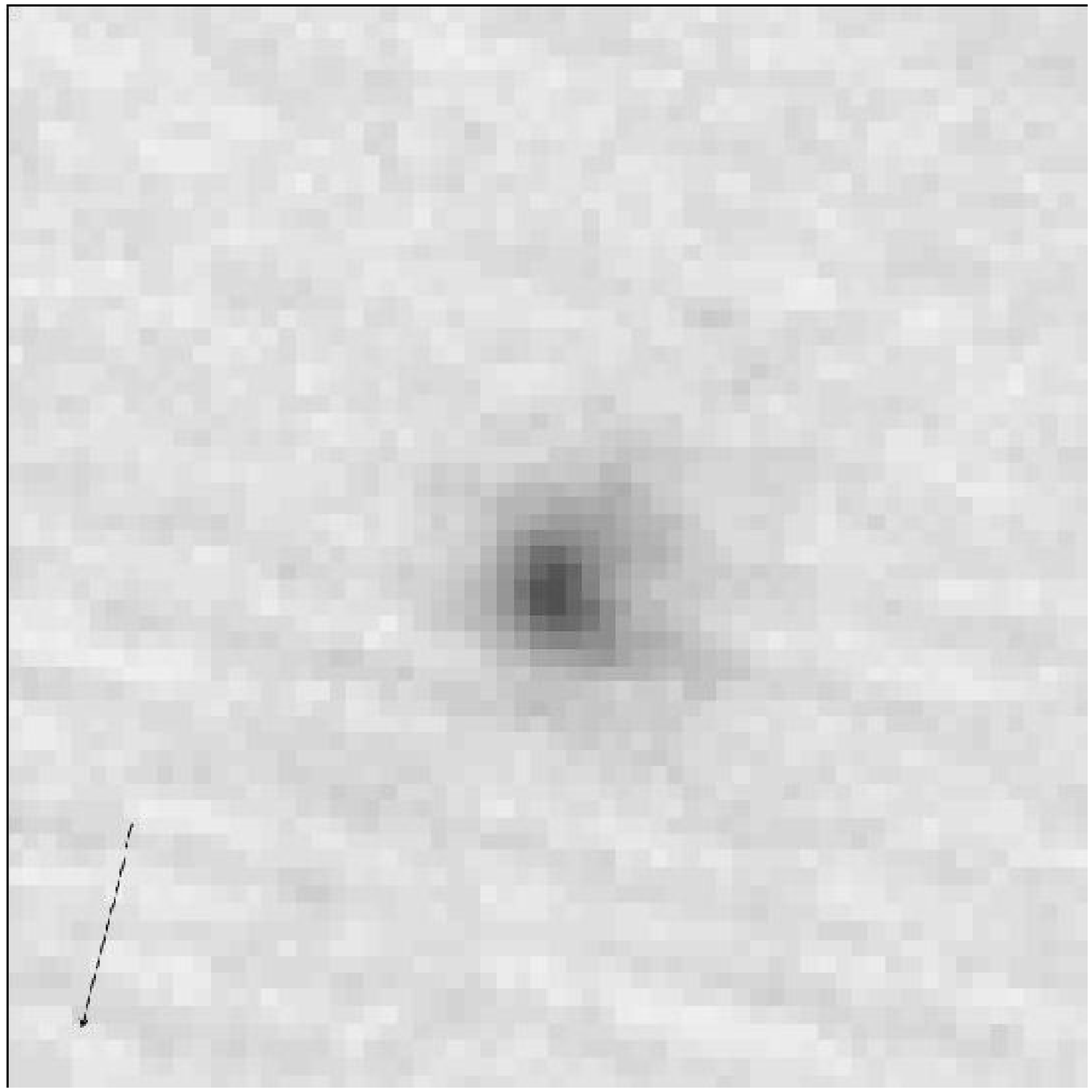}
\caption{Keck I/LWS 11.7$\mu$m image of Cygnus A. The arrow indicates north and
is 1 arcsec long.} \end{figure}
\clearpage

\begin{figure}
\plotone{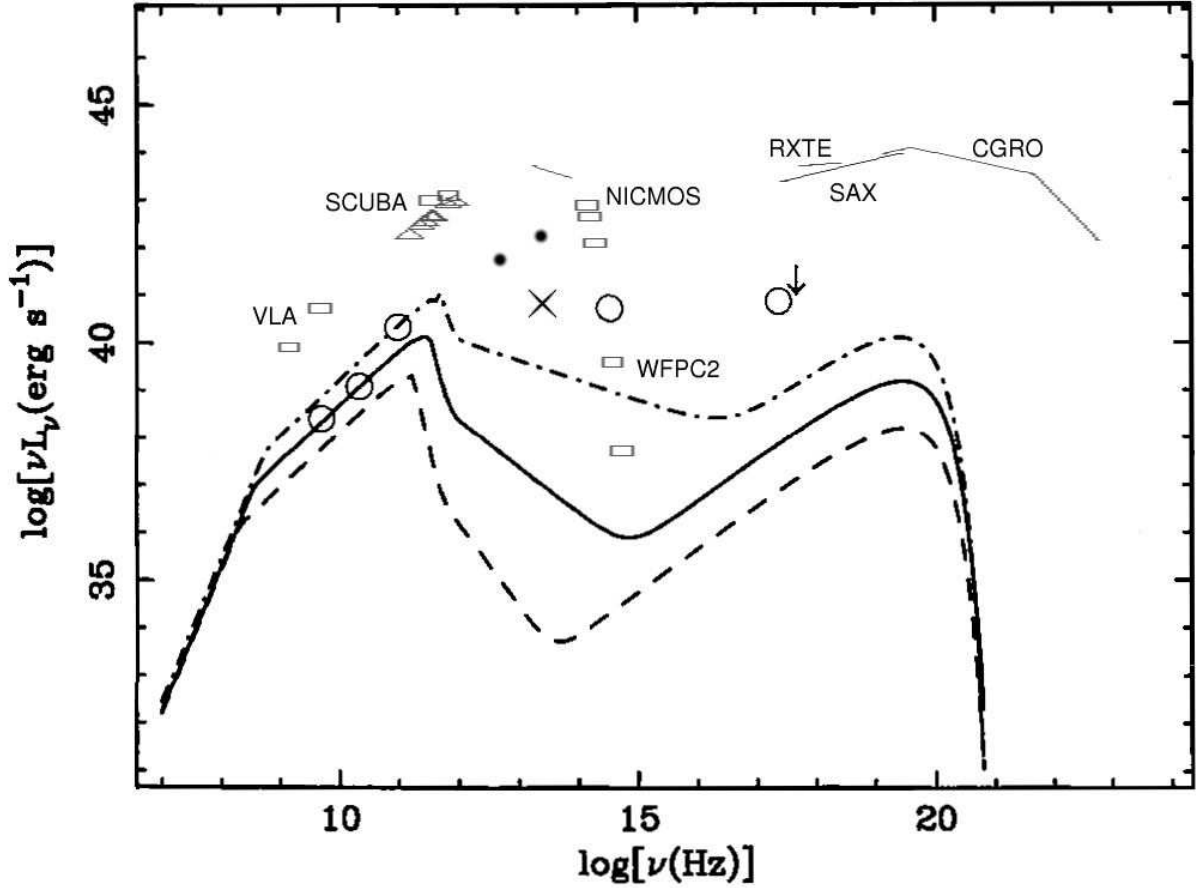}
\caption{Spectral energy distribution for Cen A (triangles, rectangles, and
thin lines, adopted from Chiaberge et al.\ 2001), and M87 (open circles), with
ADAF models for three different accretion rates (Reynolds et al.\ 1996).
Additional M87 points are from the IRAS faint source catalog (Moshir et al.\
1990) (solid circles) and Keck I/LWS (cross). The Cen A points have been
shifted up two decades for clarity.}
\end{figure}
\clearpage

\begin{figure}
\plotone{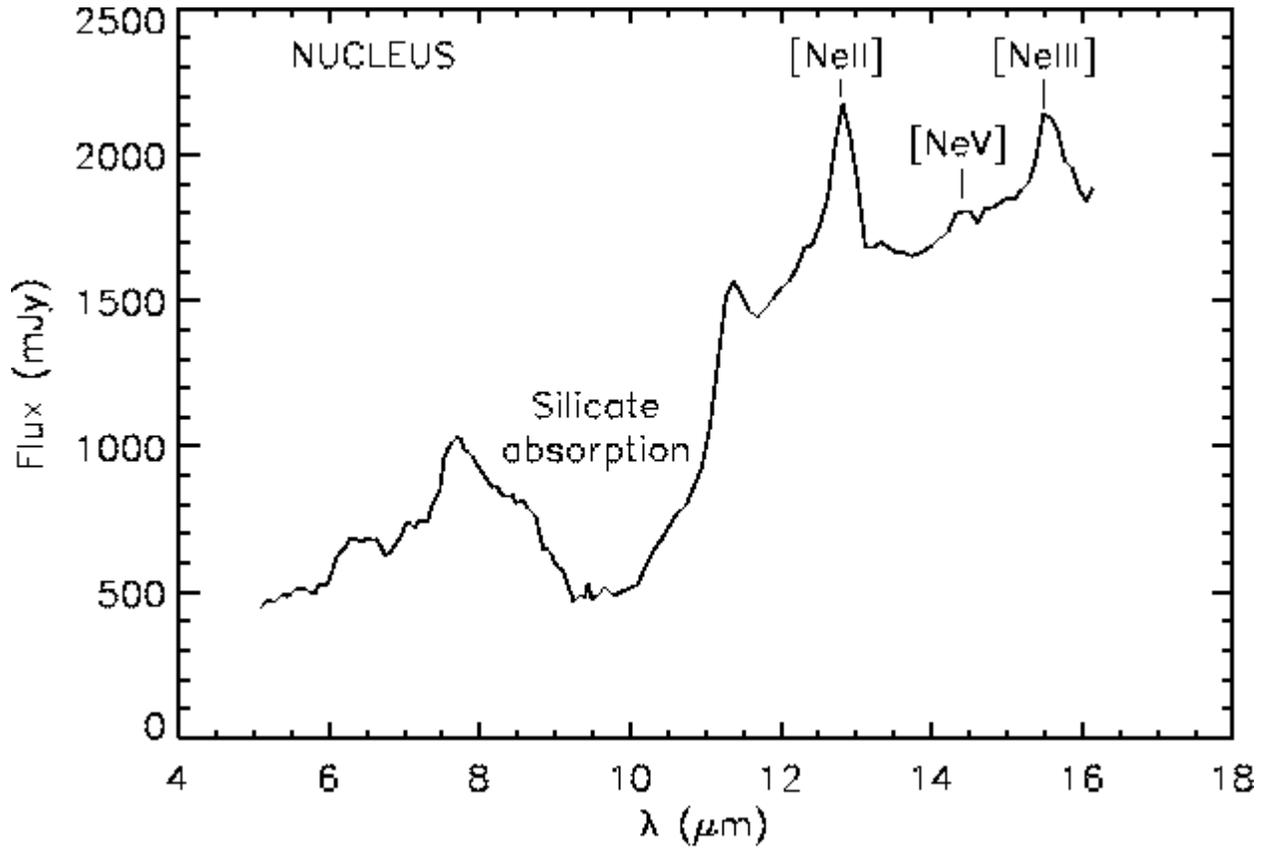}
\caption{ISOCAM CVF spectrum of Centaurus A (from Mirabel et al.\ 1999).}
\end{figure}
\clearpage

\begin{figure}
\plotone{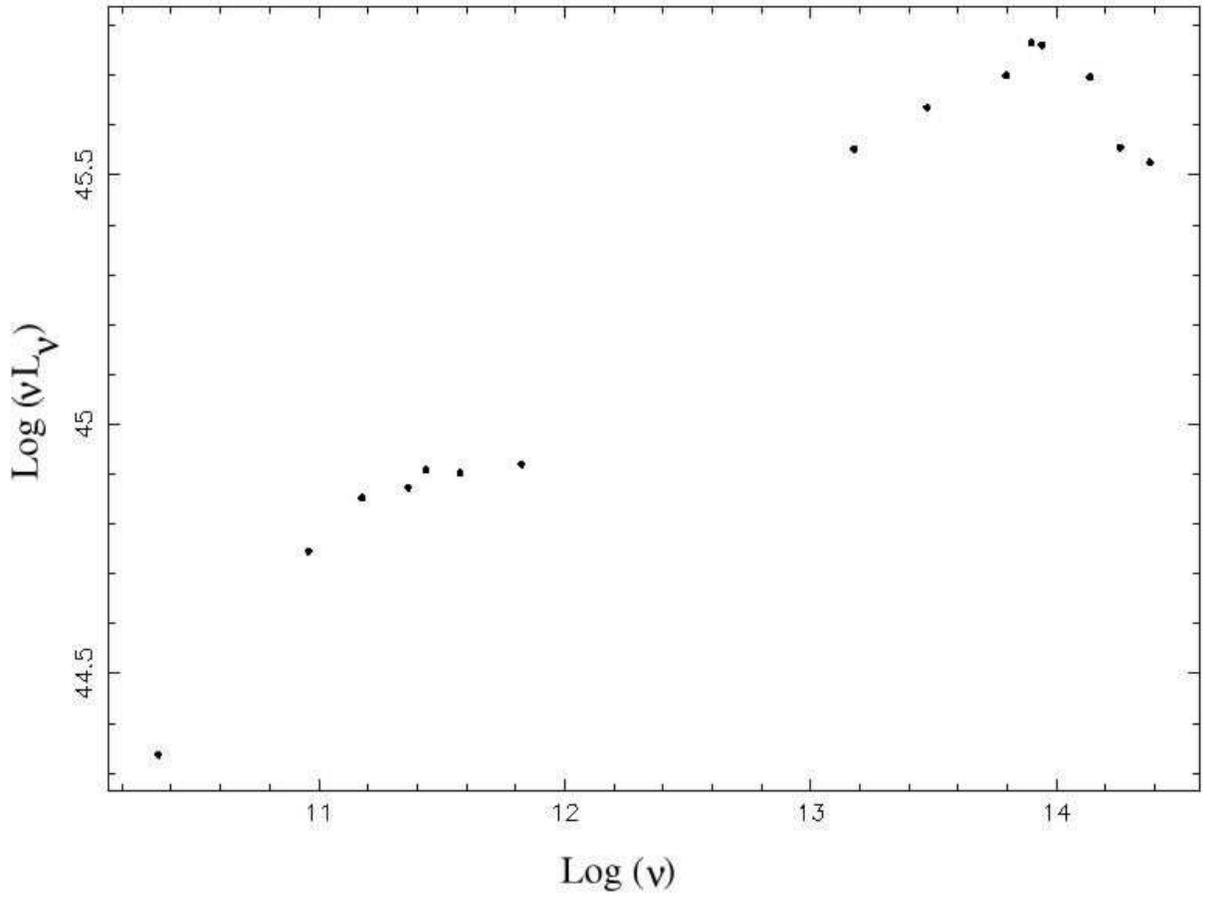}
\caption{Spectral energy distribution for 3C 273. Data were obtained from Robson et al. 1993.}
\end{figure}
\clearpage

\begin{figure}
\epsscale{0.8}
\plotone{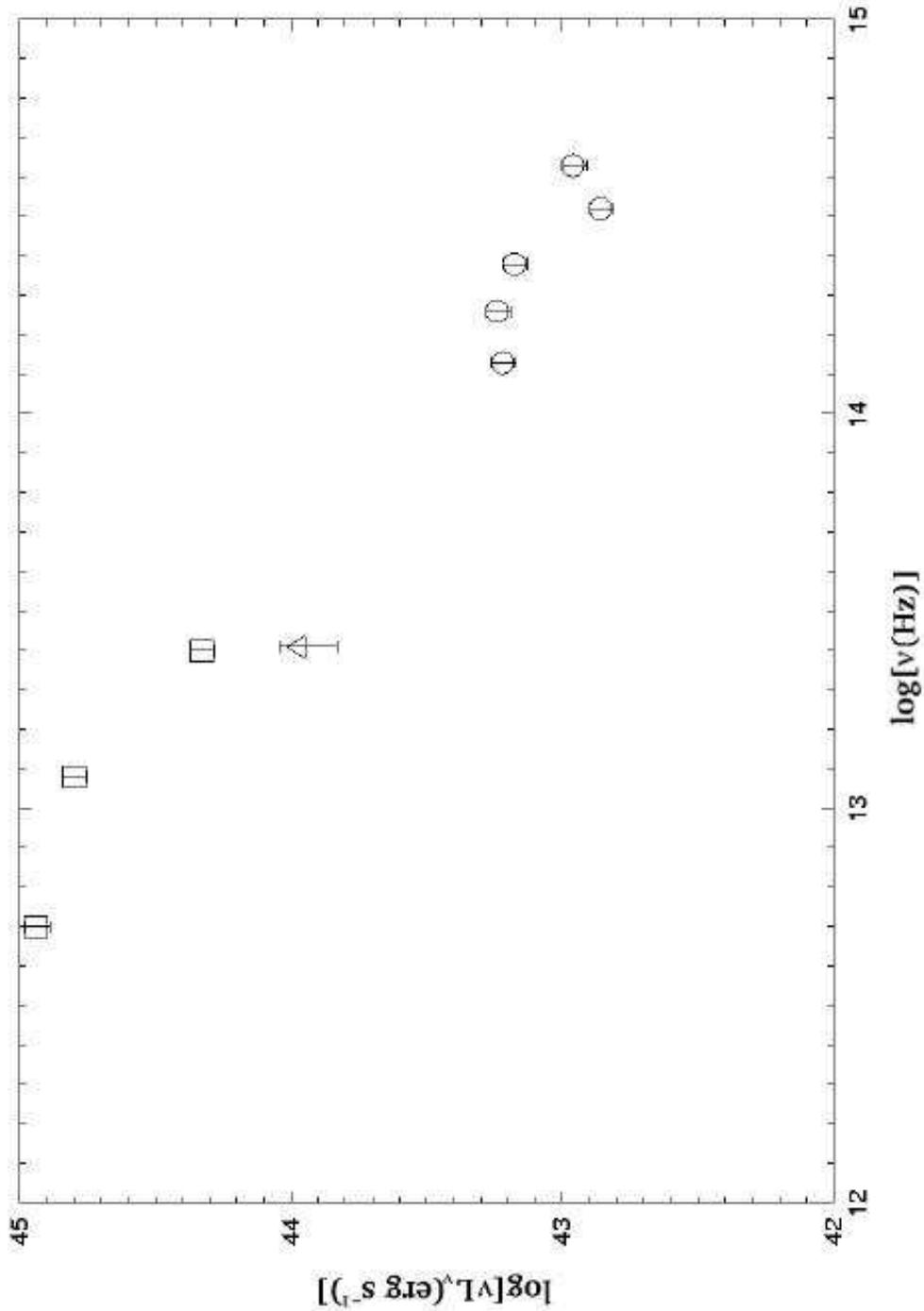}
\caption{Partial core SED for Cygnus A. Data are from IRAS (squares, Impey and
Neugebauer 1988), Keck/LWS (triangle), and Palomar 200 inch (circles, Djorgovski
et al.\ 1991). Errors for the Keck/LWS point represent uncertainty in nuclear
emission due to the extended structure. Other data are plotted with 10\% error
bars. While measurement errors were much smaller than this, we include these
larger uncertainties due to the different apertures.}
\end{figure}
\clearpage

\tablenum{1}
$$\vbox{\halign{#\hfill&\qquad#\hfill&\qquad#\hfill&\qquad#\hfill\cr
\noalign{Table 1: Flux Ratios}
\noalign{Mid-IR results are from our Keck I survey.}
Name & 11.7$\mu$m (mJy) & 1.4 GHz (mJy) & Ratio\cr
3C 47  & 10   &  3800 &   2.6E-3\cr
3C 216 & 9.7  &  4000 &   2.4E-3\cr
3C 219 & 7.0  &  8000 &   8.8E-4\cr
3C 234 & 186  &  5400 &   3.4E-2\cr
3C 382 & 27.5 &  5100 &   5.4E-3\cr
3C 452 & 22   &  10600 &  2.1E-3\cr
}}$$


\tablenum{2}
$$\vbox{\halign{#\hfill&\qquad\hfill#\cr
\noalign{Table 2: LWS photometry results for Cygnus A:}
 aperture diameter  &     flux (mJy)\cr
     (arcsec)\cr
      0.64     &              44\cr
      0.96     &              71\cr
      1.28     &              93\cr
      1.60     &             111\cr
      1.92     &             122\cr
      2.56     &             139\cr
      3.20     &             152\cr
}}$$


\tablenum{3}
$$\vbox{\halign{#\hfill&\qquad\hfill#\cr
\noalign{Table 3: IRAS photometry for Cygnus A (Impey and Neugebauer 1988):}
    12 $\mu$m &  S = 144 +/- 5 mJy\cr
    25 $\mu$m &  S = 870 +/- 5 mJy\cr
    60 $\mu$m &  S = 2908 +/- 13 mJy\cr
}}$$

\clearpage

\end{document}